# $B^2FH$, the CMB & Cosmology

G. Burbidge
University of California San Diego
La Jolla, CA

## ABSTRACT

In this talk I shall start by describing how we set about and carried out the work which led to the publication of $B^2FH$ in 1957.

I then shall try and relate this work and the circumstances that surrounded it to the larger problem of the origin and formation of the universe. Here it is necessary to look back at the way that ideas developed and how in many situations astronomers went astray. Of course this is a personal view, though I very strongly believe that if he were still here, it is the approach that Fred Hoyle would take.

I start by describing the problems originally encountered by Gamow and his associates in trying to decide where the helium was made. This leads me to a modern discussion of the origin of $^2D$, $^3He$, $^4He$ and $^7Li$, originally described by $B^2FH$ as due to the x-process. While it is generally argued, following Gamow, Alpher, and Herman, that these isotopes were synthesized in a big bang I shall show that it is equally likely that these isotopes were made in active galactic nuclei, as was the cosmic microwave background (CMB), in a cyclic universe model. The key piece of observational evidence is that the amount of energy carried by the CMB, namely about $4.5 \times 10^{-13}$ erg cm$^{-3}$.

## THE CREATION OF $B^2FH$

Of the five of us, Margaret and I are the only ones left. Thus it is natural for you to ask us to tell you how the work was carried out. What I am going to try to do is first to describe how I (we) saw the project, and secondly to talk about the way in which it has impacted on modern astrophysics.

The first thing I want to say is that while Margaret and I worked jointly, what I am describing is only the way I saw us functioning (everyone of us sees things from a different standpoint).

We had returned to the UK from the US in the autumn of 1953, and I took up an appointment at the Cavendish Laboratory in Cambridge working with Martin Ryle's radio astronomy group.

We had brought back with us many high-resolution spectra of peculiar stars: Margaret's whole expertise was on stellar spectroscopy and she had obtained many spectra of stars which show abundance anomalies. We had reduced the spectra of some of them and had shown that peculiar A stars have highly anomalous abundances of many of the heavier elements, particularly the rare earths. While Margaret was reducing more spectra in Cambridge, we had developed the idea that perhaps the anomalous abundances were due to the build-up of heavier elements with magic closed shell values, by neutron capture arising from the bombardment of heavy nuclei by protons accelerated by the changing magnetic fields on the surfaces of the stars.

I gave a talk about this in Cambridge in the winter of 1953-54, and afterwards. Willy Fowler, who was a Fulbright professor there, came up to me and told me how interested he was although at that time he only worked on light elements!

This led the three of us in the succeeding weeks to start to work together on what we all now call slow neutron capture – the s-process, going on in carbon stars and the like.

Thus I started to work in this field while I was also trying to understand radio sources, and also contend with Martin Ryle, (who could be very nice, but also extremely difficult). Willy Fowler re-introduced us to Fred Hoyle whom we had originally met in Paris in 1950. Willy of course knew Fred from his earlier visits to CalTech.

We soon developed a pattern: tea in the Cavendish, followed eventually by all of us often going over to Willy's rented house where I first found out about martinis. Frequently, we went to dinner together. In the midst of all of this quite a lot of work got started.

I'm not really sure at that time that I realized that there was a major problem to be solved, - the origin of all of the elements. I had read Fred's paper of 1946, but I was abysmally unaware of Gamow's work done in the same period.

Margaret and I had attended a meeting on the abundances of the elements held at the Yerkes Observatory in 1952, and I suspect that Margaret was more aware of the importance of this than I was at that time. But I found that discussing all of the astrophysical problems, including nuclear processes, radio sources, and cosmic rays with Fred led us to all kinds of new ideas, some of which we published.

I suspect that the full impact of the work that we did on stellar nucleosynthesis didn't really sink in until I moved to UCSD in 1962, and talked at length to Maria Mayer. This led me to realize that stellar nucleosynthesis as pioneered by Fred had completely vanquished the earlier ideas of Gamow, Alpher, and Herman, except for the x-process.

But to go back to 1953-54: The progress that we made on the s-process was also being paralleled by Al Cameron quite independently at Chalk River. Willy was the first to give him credit for this. The importance of carbon, of helium burning, (Saltpeter) and α-particle capture took up much of our discussion.

What made B$^2$FH so productive was that we relied on good nuclear, experimental data (Willy's judgment on this was supreme) and good astronomical spectroscopy (from Margaret) and a large number of blackboard and back of the envelope calculations together with much solid work. At the same time Fred Hoyle and Martin Schwarzchild were working with many people on stellar evolution.

By late 1954 Willy began to see that the group should be kept together, and encouraged us (Margaret and me) to come to CalTech in 1955. After the IAU in Dublin in 1955 we came – Margaret to Kellogg using AEC Funds, and me to Carnegie (the first theoretical fellow ever, – so that Margaret could observe – maybe.) We came in September 1955 and the collaboration went on (Fred was at CalTech for long periods) – much of the work was done in a windowless office in Kellogg. We got somewhere with the e-process using my old Brunswiga calculator and worked extensively on the r-process. From the earlier work of Fred's we knew that supernovae must play a role.

For me the clincher in this area came when I noted a paper in Physics Review Letters which reported the production of Cf$^{254}$ in the Bikini Atoll explosion. The half-life for decay of Cf$^{254}$ was about 55 days, completely mirroring as we thought was the decay of the light curve of the supernova in IC 4182 as observed by Baade in 1938.

The connection was simple, and obvious, later on it was shown to be wrong! But it had a large effect, -- at least on me. Thereby hangs a tale.

I used to go and talk to Dick Feynman about our work on nucleosynthesis and even more to him about the work I was doing independently on the energetics of radio galaxies. He was marvelous to talk to, and argue with, but only if you knew your physics. In the case of Cf$^{254}$, he liked it but was skeptical for all the right reasons. I asked him if he would believe it and he said, with a large smile, only if he saw a paper about it with Bob Christy as one of the authors. Some of you know or remember Bob's conservative reputation.

Well, I succeeded. There *is* a paper reporting this whole situation by Baade, Burbidge, Christy, Fowler and Hoyle which appeared in the Publications of the Astronomical Society of the Pacific in 1956.

By the end of 1956 we had put together a large amount of material most of which now makes up B$^2$FH. But it was hard to know how to write it up in a practical way. Providentially Fred and Willy were invited to attend a Vatican Conference on Stellar Populations early in 1957. Margaret and I were thought to be too junior to be invited. But while Willy and Fred were away for approximately two or three months, we put together a draft of B$^2$FH.

In general this would have not been possible because before 10 pages on anything could be put together Willy would be all over them, writing and expanding but always changing – what anyone of us had written. Fred, like me, wasn't so worried about details, and would go along without extensive re-writing. But now, faced with a long detailed draft there was general inclination not to break it up and start again.

Thus by the summer we had an orange-aid preprint. Who would publish it and where? I tried Chandra who was editor of The Astrophysical Journal at the time. He answered and asked questions like how much of it was original, and how much a review? He dithered though it was at least 95% original. Willy was impatient. He called up his friend, Ed Condon, who was editor of Reviews of Modern Physics, who simply accepted it and published it in the late summer of 1957. (If it were fifty years later, it would have been in the hands of referees for months, if not years!)  As you know I believe that the refereeing system is completely broken.

We were successful in these endeavors because we based our work on very good experimental and observational data.  We also respected each other's expertise and abilities and didn't want to rule out any ideas.  There was no leader.  We each brought something to the table.

It is unfortunate that Willy was singled out to get a Nobel prize and Fred was left out.  I have good reason to believe that this was because, in the eyes of the Physics community, he was seen as the leader, but as I have shown there was no leader among $B^2FH$, though Willy, of course, was the outstanding experimental physicist among us.

## $B^2FH$ and Cosmology

On the general cosmological level it is impossible to isolate the work of $B^2FH$ and its extensions to cosmology in general.  J. Robert Oppenheimer often made the argument, quite unfairly and also chronologically incorrectly, that stellar nucleosynthesis was the only successful outcome of the steady-state cosmology, which he and many others, even today, felt was a failed theory.  Apparently Oppenheimer never bothered to find out that Hoyle's original paper on nucleosynthesis was published in 1946 (Hoyle 1946) and it predated the work by Hoyle, Bondi, and Gold on the steady state cosmology published in 1948 (Bondi & Gold 1948; Hoyle 1948).

As we pointed out in $B^2FH$, the only isotopes we could not account for in stellar nucleosynthesis were $^2D$, $^3He$, $^4He$ and $^7Li$.  Thus we said that they are produced by an unknown process, the x-process.

In the late 1940's Gamow and other leading physicists, Fermi, Teller, Maria Meyer, Peierls, and others all tried to build the elements in an early big bang phase.  Of course they all failed.  They could only build $^2D$, He and some $^7Li$ because there is no stable mass 5 or 8.  With the exception of Gamow and Alpher and Herman, they all dropped the subject of primordial nucleosynthesis.  But Gamow et al were fascinated by the idea of a beginning involving a hot fireball and predicted that it would continue to expand in a black body form.  They believed in it.

Thus the idea that the lightest isotopes were made in a big bang was accepted, and today is one of the pillars of the standard cosmology.  What is not pointed out is that there is no basic theory behind it.  Prior to 1945 it was always assumed that the initial ratio of baryons to photons was very large.   Gamow found that with this ratio no elements could be built, and thus he and his colleagues <u>chose</u> a ratio such that, initially the photons dominated over the baryons, $\rho_b/\rho_{ph} = 4 \times 10^{-11}$.  A value very close to this has been assumed ever since.

Gamow's original idea of invoking a big bang stemmed from the fact that he found that the comparatively large observed abundance of helium could not be made by hydrogen burning in stars over the lifetime of the universe, because the amount produced would be much too small.  At that time the value of the Hubble constant derived by Hubble and Humason (1931) was 550 km sec$^{-1}$ Mpc$^{-1}$ giving a time scale $H_o^{-1}$ of only $2 \times 10^9$ years.  In what follows we start by updating the discussion of the helium problem.

The Origin of Helium

The largest energy densities of all of the diffuse radiation and particle fields that we can detect in the solar system are (a) cosmic rays, and (b) the microwave background. They each amount to approximately 1 eV per cubic centimeter or about $4 \times 10^{-13}$ erg cm$^{-3}$. All of the other diffuse radiation fields, IR, optical (starlight), UV, X-rays and $\gamma$ rays are much less energetic. The largest amount of energy released in nuclear reactions in stars comes through the burning of hydrogen into helium which releases about $6 \times 10^{18}$ erg gm$^{-1}$. Thus it is natural to suppose that, since the steadiest and most visible energy sources in the universe appears to be stars, that they are ultimately the source of the largest of the diffuse energy fields.

(a) Observational evidence for the background radiation field

In 1926 Eddington (1926) made an order of magnitude estimate of the energy density of starlight and found a value of $7.67 \times 10^{-13}$ erg cm$^{-3}$ which if equated to a black body equilibrium distribution corresponding to a temperature of about 3°K. A modern calculation of the same kind by Pecker and Narlikar (2006) gives a temperature of 4.2°K. Of course, this is the energy density in the vicinity of our Galaxy, and at an arbitrary point in the Universe the coincidence between this and that measured temperature of the CMB of 2.726°K is just that. The energy density of diffuse starlight in the universe is only of order $10^{-14}$ erg cm$^{-3}$.

The first measurement of the CMB was made in 1941 by McKellar (1941) from the sharp interstellar absorption lines due to CN gas in the Milky Way. He showed that the rotational levels indicated that the temperature must lie in the range 1.8°K<T<3.4°K. This fundamental result predates the Penzias and Wilson (1965) by 24 years.

It is often claimed, that the abundances calculated originally by Gamow and his collaborators, and later by many others agree so well with the observed abundances, that this is proof that the big bang occurred. But this is simply not correct. The statement that the big bang theory _explains_ the observed microwave background and also _explains_ the light element abundances is to distort the meaning of words.

Explanations in science are normally to be considered like theorems in mathematics, to flow deductively from axioms and not to be restatements of axioms themselves.

Thus the radiation-dominated early universe is an axiom of big bang cosmology, and the supposed explanation of the CMB, and the light element abundances, is a restatement of that axiom. To reiterate, the baryon density and temperature relation has to be fixed suitably in order to explain the light element abundances.

Thus it must be remembered that the whole argument is based on the idea that helium <u>was</u> made by such a fireball, and much as most people want to believe it there is no independent evidence that this ever did take place.

The direct detection by Penzias and Wilson, and the observations from space starting in 1990 by Mather, Smoot and their collaborators showed that the radiation is of blackbody form with a temperature of 2.736°K. It is of some interest to point out that detailed studies of the diffuse interstellar bands (DIBs) (cf Herbig 1995) show that they are due to grains containing magnesium tetrabenzporphyrin (Mg TBP) and $H_2$ TBP. These highly stable abiotic molecules belong to a subclass of a larger category of molecules know as porphyrins and metalloporphyrins. A low concentration of pyridine within the grains with a transition window of 2175A explains the well-known UV bump in the interstellar spectrum. Johnson (2006) has shown that the blue emission spectrum of the star HD 44179 displays the fluorescence excitation spectrum of bare MgTBP. This unique spectrum measures low temperature lab data of MgTBP in the vapor phase and has led Johnson to deduce an effective grain temperature of 2.728°K based on MgTBP's longest measured vibration state of 341 GHZ.

Thus both grain temperature and radiation temperature agree very closely. The fact that the $\lambda$ 2175 feature has now been detected in some QSOs (e.g. A0 0235 + 164 (Junkkarinen et al 2004) at z = 0.524 shows that the same temperature is present in objects with significant redshifts.

The energy density of this radiation field is $aT^4 = 4.420 \times 10^{-13}$ erg cm$^{-3}$ ($a = 7.564 \times 10^{-15}$ erg cm$^{-1}$ T$^{-4}$).

(b) <u>Possible Source of the Energy in the CMB</u>

It is natural to suppose that the energy carried by the CMB is most likely to have been generated by hydrogen burning in stars, since this is the most effective process of conversion of mass to energy involving a set of nuclear reactions. To see whether or not this is technically possible we need to know:
(a) The energy density of the radiation (given above)
(b) The mass ratio, $^4$He/total mass
(c) The timescale over which hydrogen burning has taken place, i.e., how many hydrogen burning stars are involved and for how long must the process have been going on.

The mass fraction of $^4$He is measured by obtaining the number density $^4$He/H from optical or radio spectroscopy of hot stars or gaseous nebulae.

For a variety of hot stars and nebulae analyses of the optical recombination spectra have led to the conclusion that n($^4$He)/n(H) is about 0.1 corresponding to a mass fraction lying in the range 0.23 – 0.25. The radio recombination lines have been used to determine the $^4$He/H ratio and the value obtained is somewhat lower than that found by the optical methods.

What is required if we try to attribute all of the $^4$He to hydrogen burning? For a flat universe with the cosmological constant $\Lambda$ = 0, the mean density is $\zeta_c = \frac{3H_o^2}{8\Pi G}$ for $H_o$ = 60km sec$^{-1}$ Mpc$^{-1}$, $\rho_c = 7.2 \times 10^{-30}$ gm/cm$^3$.

If the $^4$He is universally 24% by mass, the energy realized in burning hydrogen is $1.728 \times 10^{-30} \times 6 \times 10^{18}$ erg/cm$^3$ = $1 \times 10^{-11}$ erg/cm$^3$.

But this is an estimate based on a simple theoretical model and there is no observational basis for it. The density of baryonic matter both seen and unseen in galaxies, which is all that we know based on observations, gives a mean density of $3 – 7 \times 10^{-31}$ gm cm$^{-3}$. Such values have been obtained by counting galaxies and using observed mass–to-light ratios starting with Oort, Kiang, and others. Thus if we take $\rho_b = 5 \times 10^{-31}$ gm cm$^3$, the energy released in converting 24% of the total mass to $^4$He is $7.2 \times 10^{-13}$ erg cm$^{-3}$. If this energy is degraded to black body energy, at the temperature will be about 3.1°K.

This is so close to the <u>measured</u> temperature of the CMB that it strongly supports the idea that the observed CMB and the production of $^4$He are closely related and that the conversion has come from hydrogen burning in stars. However, there is a serious problem associated with the time scale for this process. The characteristic time scale for the expanding universe is given by $(H_o)^{-1}$, and for $H_o = 60$ km sec$^{-1}$ Mpc$^{-1}$ this "age" is $5 \times 10^{17}$ sec $= 17 \times 10^9$ years. For "average" solar type main sequence stars the luminosity is $= 4 \times 10^{33}$ erg sec$^{-1}$ and thus the rate of conversion about 2 erg sec$^{-1}$.

A typical galaxy with $M_B = -20$ has a luminosity of $5.6 \times 10^{43}$ erg sec. Thus in $17 \times 10^9$ years it will synthesize about $10^8$ $M_\odot$ of He$^4$, so that the $M_{He}/M \simeq 10^{-3}$, far below the measured fractorial abundance.

Historically it is this argument which led Gamow (1946) to the proposal that helium must have come from the early universe. As was stated earlier in 1946 when he was originally working on the problem, Gamow had to use for the time scale a value for $H_o$ of 550 km sec$^{-1}$ Mpc$^{-1}$ (Hubble and Humason 1936) giving $H_o^{-1} \sim 2 \times 10^9$ years, a factor of 10 less than the real expansion rate.

As we have shown above, the longer time scale given by a more realistic value of $M_o$ does not work either. Gamow immediately gave up on the scheme and concluded that helium (and the other elements) must have been made in an early universe. Thus he and his colleagues Alpher and Herman chose parameters for an early universe that would solve the problem. Since there is no theory from which the initial condition can be predicted they had to <u>choose</u> an initial ratio of baryons to photons of $\sim 4 \times 10^{-11}$.

It was then possible to explain the observed abundances of $^4$He, and later Hoyle and Tayler (1964) and Peebles (1966) made more accurate calculations. The temperature of the hot fireball cannot be predicted (cf Turner 1993) but Gamow and his colleagues and successors guessed that it might be 5°K or even 10°K at different times. For such temperatures the energy density of the black body radiation is 8 to 120 times greater than the observed value. Of course Gamow and his colleagues had completely overlooked the 1941 measurement of McKellar.

As I pointed out earlier, while $^4$He could be produced in an early universe, primordial nucleosynthesis could not build the elements beyond $^7$Li.

However, Gamow and his colleagues, and much later Dicke and others who reworked the problem, took the position that this had led to the most likely cosmological model, and with the direct detection of the CMB by Penzias and Wilson in 1965 the belief in the correctness of this cosmological model became very strong. Detailed calculations were made by several groups (following Hoyle and Tayler and Peebles) of the abundances of the light isotopes D, $^3$He, $^4$He and $^7$Li which could be made in an early universe. It is frequently claimed that to detect and measure the abundances of these light isotopes, particularly the D/H ratio, and get agreement with the calculated values is in some sense is proof of the correctness of the model. This is not correct, since the initial conditions assumed by theorists, near to t = 0, were <u>chosen</u> to make the model work.

However, the fact the observed CMB temperature gives an energy density very close to that expected if the observed helium mass fraction is due to hydrogen burning is so remarkable that it suggests that there must be a direct connection.

We can see now that there are actually three possibilities:

(1) Most of the helium was made in a big bang, and the parameters required are those chosen in the conventional model. This is the most popular view but in its present form it requires that we <u>choose</u> an initial photon/baryon ratio, invoke a "magical" inflation era, and assume the presence of a large amount of dark non-baryonic matter, and dark energy (creation energy). These are four assumptions for which we have no basic theory, nor direct observational evidence. Just authoritarian belief.

(2) If most of the helium was produced by hydrogen burning in main sequence stars, this must have taken place over a much longer timescale than the classical time $H_o^{-1}$. The timescale needs to be $10^{11} - 10^{12}$ years and thus an alternative cosmological model is indicated.

(3) A cosmological model with a long timescale ($10^{11}$-$10^{12}$ years) in which there are distributed creation sources (little big bangs, from supermassive objects) in which the observed abundances of $^4$He, $^3$He, D and $^7$Li, are produced can do this.

In my view it is the historical steps which I outlined earlier which led to the strongly held belief that only the first option listed above is likely to be correct. The only way to avoid this is to go to a cyclic universe model in which the timescale is infinite. Within the framework of such a model either (2) or (3) is viable.

Such a cyclic model was first proposed by Hoyle, Burbidge & Narlikar in the early 1990s (cf Hoyle et al 2000). It is a modified steady state model.[1] Rather than taking place in an instant, as is the case in any evolutionary cosmological model, the creation processes take place in the centers of active galaxies over a long time scale. As is the case in the big bang we do not have a detailed understanding of the creation process, though the classical theoretical approach is based on the C field theory which Hoyle and Narlikar developed in the 1960s.

What we do have is much observational evidence of intense activity in the centers of many comparatively nearby low redshift galaxies. It is generally agreed that this activity is driven by the presence of supermassive black holes in the centers of the

---

[1] In this model the universe oscillates (expands and contracts over a time scale ~ 50 G years within a slow overall expansion time scale of $10^{12}$ years. Since we are currently in the expansion phase of the cycle it is reasonable to ask why the temperature and energy density of the CMB remains constant. As was shown by Hoyle et al (2000) the scale factor S(t) is given by S(t)=exp t/P x F (cos 2 Π t/Q). P and Q are constants of the model and F(cos2Πt/Q) is oscillatory with respect to COS 2 Π t/Q. It is assumed that P>>Q.

If there were no additions from cycle to cycle the CMB would decline as cxp(-4t/P). Thus an addition is needed in each cycle to maintain the measured energy density. Thus for P /Q in the range 10 – 20 the addition of new background radiation per cycle is about 5 x $10^{-14}$ erg cm$^{-3}$ which we believe comes from the thermalization of starlight occurring near the minimum phase of the cycle.

It is shown in Hoyle et al (2000, Chapter 16) that the radiation (starlight) which has been thermalized has been traveling for ~ $10^{11}$ years, and the microwaves even more. The essential point is that the radiation which is being produced and mixed in has come from a very large distance ~$10^{29}$ cm and it does so with the intergalactic particle density very low, permitting the radiation to travel freely.

galaxies. The conventional view is that the energy is gravitational energy released by the accretion of matter from the surrounding stars, but we believe that the efficiency of this process is much too low to explain what we see. Thus, we believe that the creation processes are occurring in the very strong gravitational fields very near to the black holes. It is the energy emitted in this process which not only gives rise to new galaxies, but which, can overcome the deceleration due to gravity, giving rise to acceleration of the expansion (if this effect is real). Eventually the galaxies will be far enough apart so that the pressure exerted by the active systems will not be enough to overcome the gravitational deceleration. At this stage the expansion will cease, and the universe will start to contract. However, long before it can collapse to very small dimensions, and before the galaxies would overlap, the pressure and activity of the creation process will begin to dominate, and the universe will re-expand. This cyclic process can continue indefinitely, with a cycle time of order 40 - 50 G years.

Thus in several cycles we have a long enough time scale for processes (2) or (3) to build the helium and the other light isotopes. For this model we predict that a large fraction of the matter is dark baryonic matter, largely made up of evolved stars (white dwarfs, neutron stars and black holes) but there is no need to invoked the existence of non-baryonic matter.

(a) <u>Explosive Processes in Galactic Nuclei</u>

Starting with the study of the physics of the collapses of supermassive stars by Hoyle & Fowler and the detailed calculations of Wagoner, Fowler & Hoyle (1967) it became clear that the lightest isotopes could also be made in a very short timescale in an explosive process at very high densities, and not necessarily only in the initial stages of a big bang.

As we described earlier, in the big bang the conditions for primordial nucleosynthesis are that $\rho_b \simeq 10^{-5}$ gm cm$^{-3}$, $T = 10^9$ K and $t = 10^2$ secs. This corresponds to a baryon/photon ratio $\sim 10^{-11}$. As was shown in Chapter 10 of Hoyle et al (2000) the light isotopes can also be synthesized in a situation in which $\rho_b \approx 10^9$ gm cm$^{-3}$, $T = 10 \times 10^9$ K, and $t = 10^{-13}$ sec. In this case the baryon/photon ratio is of order unity.

In a cyclic universe these latter conditions may be reached in the nuclei of active galaxies in regions close to supermassive black holes, where creation takes place.

Thus synthesis of helium and the other light isotopes may well occur in "little big bangs" distributed in space and time through the cycles.

(b) <u>Synthesis Through Hydrogen Burning in Stars</u>

In this situation, the much longer time scale associated with the cyclic universe allows the observed abundance of $^4$He to be built up over $\sim 10^{11}$ years, or several cycles in the model.

Of course deuterium and $^3$He are also built, but deuterium is largely destroyed. Stars of mass 1 M☉ to 2M☉ are particularly efficient at producing $^3$He which is then mixed into the convective envelopes of these stars when they reach the giant branch and are then distributed into the interstellar gas. Thus it is hard to understand the very low ratio of $^3$He to $^4$He which is observed ($\sim 2 \times 10^{-5}$). For stars of larger mass the convective core is large enough so that much of the $^3$He is convected inward and burnt.

There are two possible ways out of this problem. The first is to argue that most of the hydrogen burning takes place in stars with M > 2 M☉ i.e. most stellar nucleosynthesis which gives rise to the observed abundance of $^4$He has occurred in more massive stars.

Alternatively it has recently been suggested by Eggleton et al (2006) that mixing takes place in the supposedly stable zone between the hydrogen burning shell and the base of the convective envelope so that very little of the $^3$He can survive to escape from the surface.

Deuterium produced in the interior clearly cannot survive. However, we know that it is produced in stellar flares, and Burbidge & Hoyle (1998) have argued that the typical abundance ratio D/H ~ $10^{-5}$ arises in this way. There is a considerable amount of observational evidence concerning the D/H ratio in our Galaxy. For example values based on ultraviolet in spectra of hot stars gives D/H = 2 – 1.5 x $10^{-5}$, while data obtained using the 327 MHz radio line in emission from interstellar gas in the direction of the galactic anti-center give a ratio of $2.3 ^{+1.5}_{-1.3}$ x $10^{-5}$.

Thus it is not yet clear whether $^4$He and the other light isotopes are produced in little big bangs, or in hydrogen burning in stars.

A possible difficulty with stellar nucleosynthesis is that we must also reconcile this model with the fact that our galaxy contains old stars with apparently normal helium abundances but very small abundances of the heavier elements. For example, values of Fe/H lower than the solar value by factors up to $10^5$ have been found.

On the other hand there are some stars in which it has been found that $^3$He and $^4$He are comparable in abundance, for example 21 Aquilae (Burbidge & Burbidge, 1957), and this has to be a result of nuclear processes in the stars.

**CONCLUSION**

B$^2$FH and Cameron in 1957 were able to explain how all of the isotopes in the periodic table with the exception of $^2$D, $^3$He, $^4$He and $^7$Li could have been synthesized in stars.

In the cyclic (quasi-steady state) cosmological model, the long time scale means that the other light isotopes, and particularly the high abundance of helium, could have been synthesized as a result of creation in the centers of active galaxies. It may have been made directly in the creation processes going on in the center of many galaxies, or by hydrogen burning in massive stars, later in the evolution of galaxies.

Thus Oppenheimer's cynical view of the steady state cosmology can be stood on its head. From our standpoint, the observational data, and in particular the energy which must been released in the burning of hydrogen to produce helium, suggests that it has given rise to the <u>observed</u> microwave background. This release has taken place over a long period, too long for a big bang universe. Thus the observational data favor a cyclic universe model, in which <u>all</u> of the isotopes were made.